\documentclass[namedreferences]{SolarPhysics}
\usepackage[optionalrh]{spr-sola-addons} 
\usepackage{graphicx}        
\usepackage{color}           
\usepackage{url}             
\usepackage{enumerate}

\newcommand{\aap}{    {\it Astron. Astrophys.}}

\newcommand{\apj}{    {\it Astrophys. J.}}

\newcommand{\solphys}{{\it Solar Phys.}}

\newcommand{\ssr}{    {\it Space Sci. Rev.}}

\begin{document}

\begin{article}

\begin{opening}

\title{Investigation of the X-ray Emission of the Large Arcade Flare of 2 March 1993}

%
\author{J.~\surname{Jakimiec}\sep
        M.~\surname{Tomczak}
       }

%
\runningauthor{J.\,Jakimiec \& M.\,Tomczak} \runningtitle{X-ray Emission of the Large Arcade Flare}

%
  \institute{Astronomical Institute, University of Wroc{\l }aw,
  ul. Kopernika 11, 51-622 Wroc{\l }aw, Poland,
                     email: \url{jjakim; tomczak@astro.uni.wroc.pl}\\
             }

\begin{abstract}
A large arcade flare of 2 March 1993 has been investigated using X-ray observations recorded by the {\sl Yohkoh} and GOES satellites and the {\sl Compton Gamma Ray Observatory}. We analyzed quasi-periodicity of the hard-X-ray (HXR) pulses in the flare impulsive phase and found close similarity between the quasi-periodic sequence of the pulses with that observed in another large arcade flare of 2 November 1991. This similarity helped to explain the strong HXR pulses which were recorded at the end of the impulsive phase, as due to an inflow of dense plasma (coming from the chromospheric evaporation) into the acceleration volume inside the cusp. In HXR images a high flaring loop was seen with a triangular cusp structure at the top, where the electrons were efficiently accelerated. The sequence of HXR images allowed us to investigate complicated changes in the precipitation of the accelerated electrons toward the flare footpoints. We have shown that all these impulsive-phase observations can be easily explained in terms of the model of electron acceleration in oscillating magnetic traps located within the cusp structure.
Some soft-X-ray (SXR) images were available for the late decay phase. They show a long arcade of SXR loops. Important information about the evolution of the flare during the slow decay phase is contained in the time variation of the temperature, $T(t)$, and emission measure, EM$(t)$. This information is the following: i) weak heating occurs during the slow decay phase and it slowly decreases; ii) the decrease in the heating determines slow and smooth decrease in EM; iii) the coupling between the heating and the amount of the hot plasma makes the flare evolve along a sequence of quasi-steady states during the slow decay phase (QSS evolution).
\end{abstract}
%
\keywords{Flares, Energetic Particles, Impulsive Phase, Oscillations}

\end{opening}

\section{Introduction}
     \label{intr}

\begin{figure}
\centerline{\includegraphics[width=0.8\textwidth]{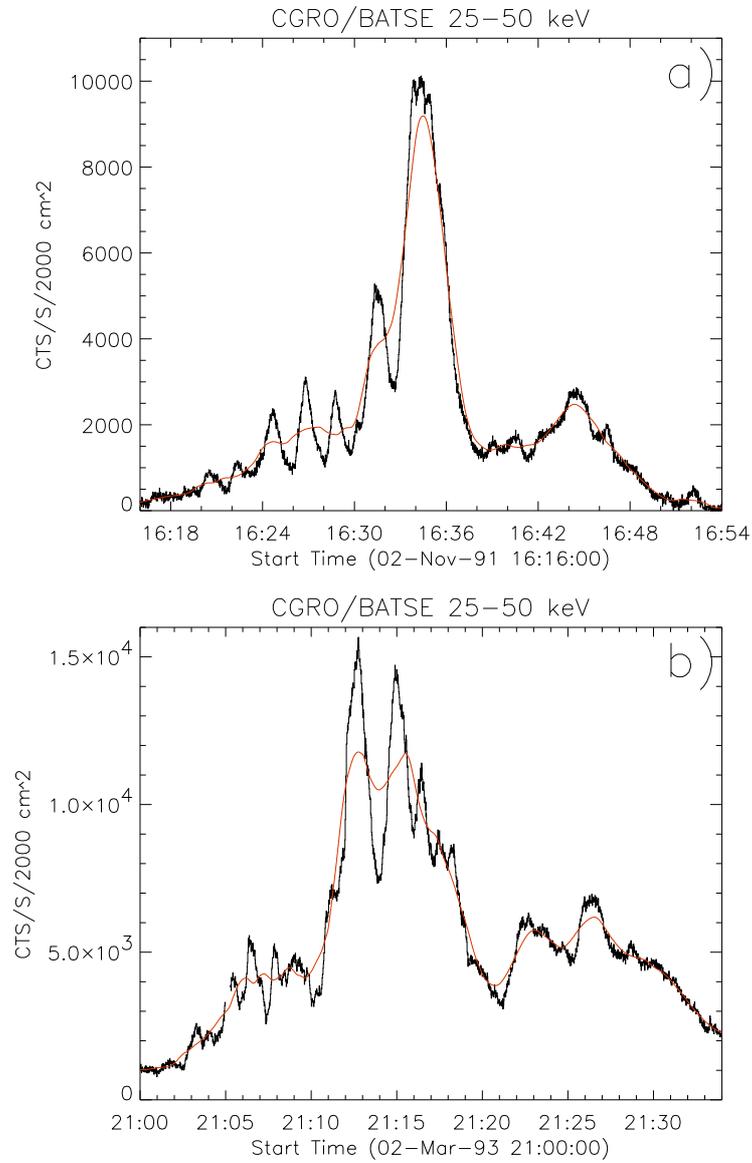}}
\hfill \caption{The CGRO/BATSE HXR light curves for two flares: a) 2 November 1991, b) 2 March 1993. The red curves show the running mean of the original light curves, see the text for discussion.} \label{fig1}
\end{figure}

Quasi-periodic variations were observed in the hard X-ray (HXR) emission of many flares (\opencite{lip78}; see also the review of \inlinecite{n+m09} and references therein). In our previous papers (\opencite{paper1} (Paper I), \citeyear{paper2} (Paper II)) we investigated quasi-periodic oscillations in flares with periods $P = 10-60$~s, but in Paper I we have found three flares with periods $P > 120$~s. They turned out to be large arcade flares. Investigation of the quasi-periodic oscillations in such large flares is very important, since their large sizes allow us to investigate the structure of the oscillation volume more comprehensively. Unfortunately, appropriate observations of the X-ray oscillations in such large flares are very rare. In Paper III \cite{paper3} we investigated such a large arcade flare of 2 November 1991. Its HXR light curve is shown in Figure~\ref{fig1}a. We have found there a direct observational evidence that the strong HXR pulse at 16:34-16:35 UT is the result of dense plasma coming from the chromospheric evaporation and flowing into the acceleration volume located within the loop-top cusp structure.

In Paper III we have also found that
\begin{enumerate}[i)]
\item the precipitation of accelerated electrons from the cusp structure is strongly asymmetric, {\it i.e.} there is strong difference in the precipitation into the northern and southern arms (``legs'') of the flaring loop;
\item there are significant changes of the precipitation with time;
\item the properties of precipitation depend on the energy of accelerated electrons.
\end{enumerate}
It has been shown that these complicated properties of the precipitation can be easily explained in terms of our model of oscillating magnetic traps (see Paper III).

In the present paper we investigate a large arcade flare of 2 March 1993. Its HXR light curve is shown in Figure~\ref{fig1}b. We see close similarity of the light curves in Figures~\ref{fig1}a and \ref{fig1}b which indicates that there is close similarity in the impulsive-phase development of these two flares. Section~\ref{sec2} contains the analysis of observations, Section~\ref{sec3} presents the discussion of the decay phase of the flare, and summary of the paper is given in Section~\ref{sec4}.

\section{Observations and Their Analysis} \label{sec2}

\begin{figure}
\centerline{\includegraphics[width=1\textwidth]{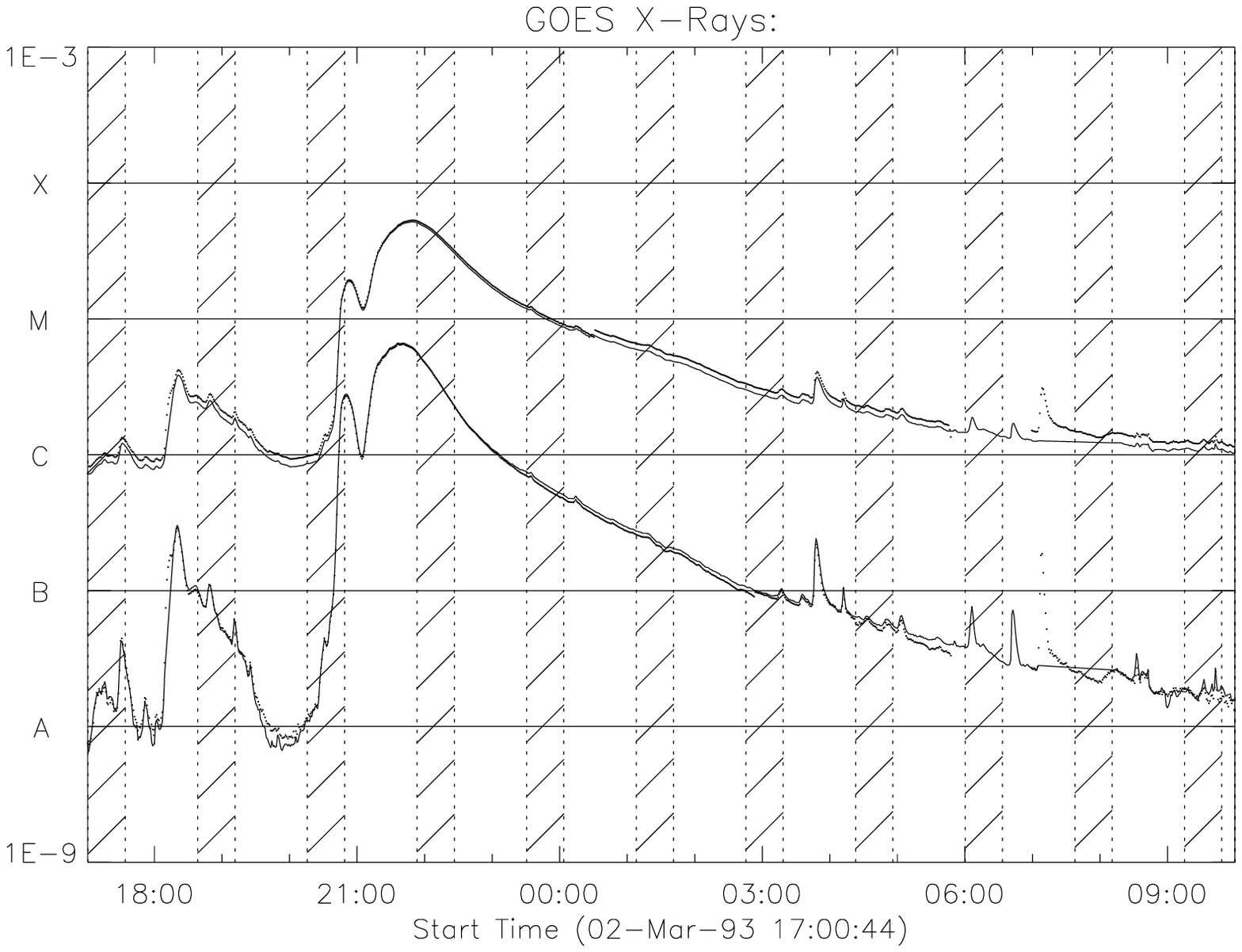}}
\hfill \caption{The standard GOES soft X-ray light curves for the flare of 2 March 1993 (upper curve -- 1-8 \AA\ range, lower curve -- 0.5-4 \AA\ range, records from two GOES satellites are displayed). The hatched areas show the {\sl Yohkoh} satellite nights.} \label{fig2}
\end{figure}

\begin{figure}
\centerline{\includegraphics[width=0.75\textwidth]{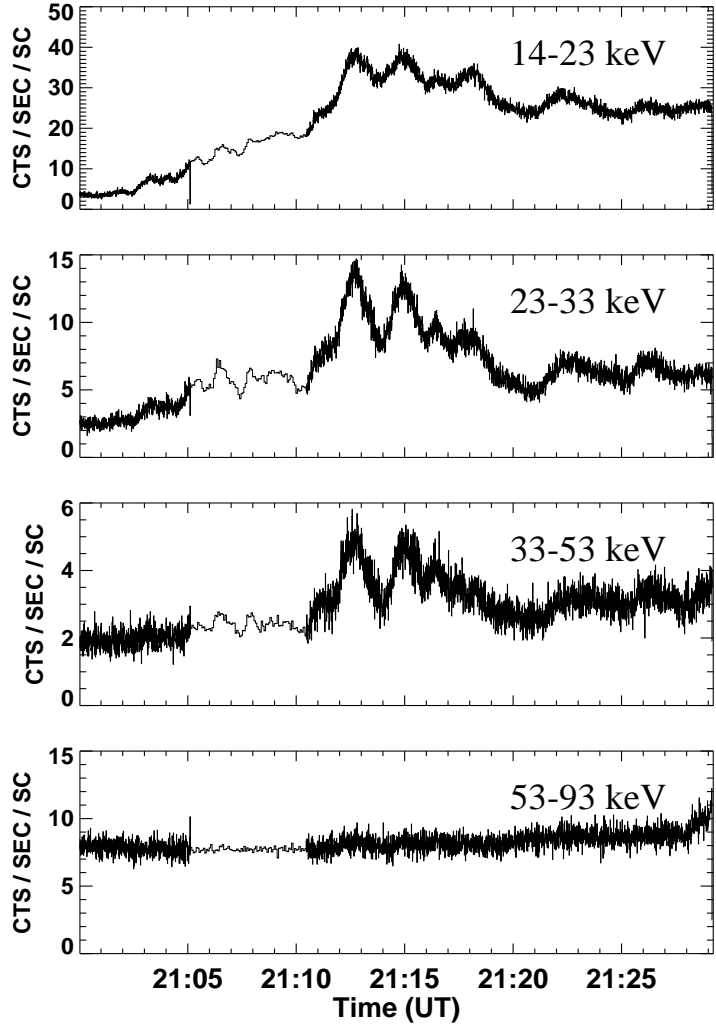}}
\hfill \caption{The {\sl Yohkoh} hard X-ray (HXR) light curves in four energy ranges.} \label{fig3}
\end{figure}

In the present paper we investigate a large arcade flare which occurred at the eastern limb on 2 March 1993. It was a long duration event (LDE) of GOES class M5.1 (see Figure~\ref{fig2}). The HXR light curves, recorded by the {\sl Yohkoh Hard X-ray Telescope} (HXT; \opencite{kos91}) and the {\sl Compton Gamma Ray Observatory Burst and Transient Source Experiment} (CGRO/BATSE; \opencite{fis92}) are shown in Figures \ref{fig3} and \ref{fig4}a, respectively. The nominal energy range of the BATSE observations is $h{\nu} > 25$~keV, but in Paper III we have found that actual BATSE energy range was $h{\nu} > 33$~keV. The HXRs began to rise at 21:01 UT (beginning of the impulsive phase). Unfortunately, there were no {\sl Yohkoh} soft-X-ray (SXR) observations for the flare impulsive phase. Some SXR images were available only for the late phase of the flare decay (see Section~\ref{sec23}).

\subsection{Analysis of Quasi-periodicity of the HXR Pulses}\label{sec21}

\begin{figure}
\centerline{\includegraphics[width=1\textwidth]{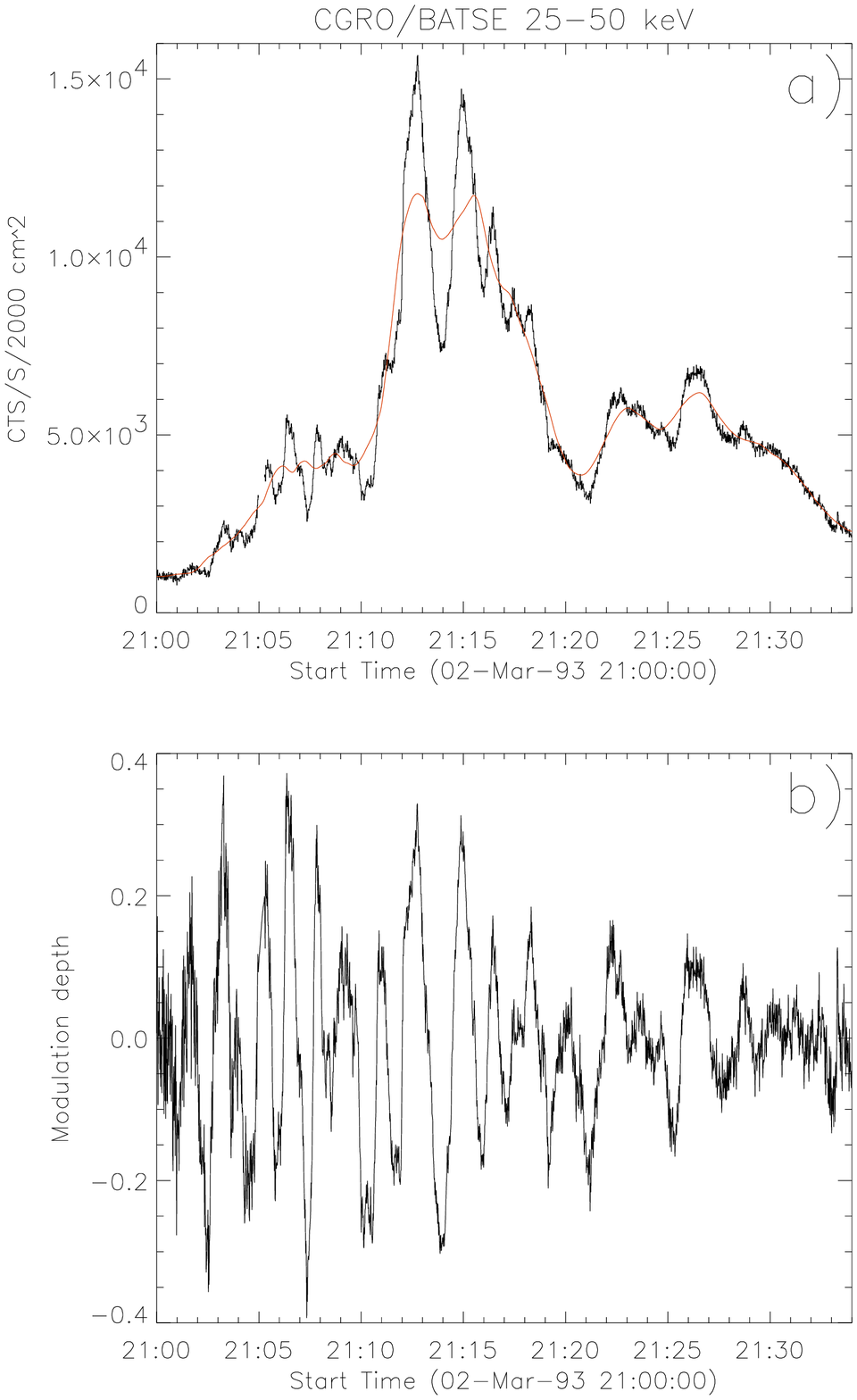}}
\hfill \caption{a) The CGRO/BATSE HXR light curve. The red, smoother, curve is the running mean of the original light curve. b) The normalized light curve, $S(t)$ [see Equation~(1)].} \label{fig4}
\end{figure}

We have investigated the quasi-periodicity of the HXR pulses using our standard method described in Papers II and III. We have calculated the normalized time series:
\begin{equation}
S(t) = \frac{F(t) - \hat{F}(t)}{\hat{F}(t)},
\end{equation}
where $F(t)$ is the measured HXR flux and $\hat{F}(t)$ is the running average of $F(t)$. The red curve in Figure \ref{fig4}a shows $\hat{F}(t)$ calculated with averaging time $\delta{t} = 120$\,s. The normalized time series, $S(t)$, is shown in Figure \ref{fig4}b. Next we have measured time intervals, $P_i$, between successive HXR peaks and calculated the period, $P=\langle P_i \rangle$, and its standard (r.m.s.) deviation, ${\sigma}(P)$. Our criterion of a quasi-periodicity is ${\sigma}(P)/P \ll 1$. The values of $P$ and ${\sigma}(P)$ are given in Table~\ref{tab1}.

\begin{table}
 \caption{Parameters of the quasi-periodic oscillations for the 2 March 1993 flare}
 \label{tab1}
\begin{tabular}{cccc}
\hline \multicolumn{2}{c}{During impulsive phase rise} & \multicolumn{2}{c}{At the maximum of impulsive phase} \\
\multicolumn{2}{c}{21:01--21:11 UT} & \multicolumn{2}{c}{21:12--21:16 UT} \\
 \cline{1-2}  \cline{3-4} \\
 $P$ [s] & Amp $S$ & $P$ [s] & Amp $S$ \\
 & & & \\
 \hline
94$\pm$8 & 0.49 & 134 & 0.56 \\
 \hline
\end{tabular}
\begin{list}{}{}
\item $P$ is the mean time-interval between successive HXR peaks, and \\ Amp $S$ is the mean value of full amplitude of function $S(t)$ \\ measured for individual HXR peaks.
\end{list}
\end{table}

Figures \ref{fig3} and \ref{fig4}, together with Table~\ref{tab1}, show clear quasi-periodicity of the HXR pulses during the impulsive phase rise (21:01-21:12 UT). Table~\ref{tab1} also shows that the period $P$ is longer near the impulsive phase maximum than during the impulsive phase rise. We explain this effect as being due to quick increase in the density inside the acceleration volume, which causes significant decrease in the Alfv$\rm{\acute{e}}$n speed, $v_{\rm A}$ (see Section~\ref{sec3}).

The light curves in Figures \ref{fig3} and \ref{fig4}a can be divided into two components: the HXR pulses and a ``quasi-smooth'' component (emission buried beneath the pulses in the figures). The full amplitude, Amp $S$, of function $S(t)$ [Equation~(1)] for a HXR pulse is a measure of the ratio of the pulse intensity to the quasi-smooth component; the mean values of Amp $S$ are given in Table~1. Our interpretation of the quasi-smooth component has been given in Paper II, where we explained it as a result of superposition of the emission generated by many magnetic traps whose oscillations are shifted in phase.

\subsection{Investigation of the HXR Images} \label{sec22}

The sensitivity of the {\sl Yohkoh} HXT was moderate and the counting rates were low in this flare (see Figure~\ref{fig3}); therefore it was necessary to apply rather long integration time in the reconstruction of HXR images.

\begin{figure}
\centerline{\includegraphics[width=0.8\textwidth]{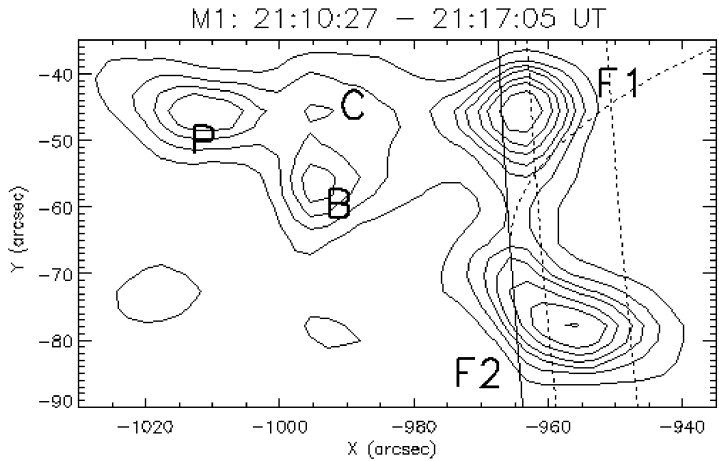}}
\hfill \caption{The {\sl Yohkoh} HXT 23-33 keV image for the HXR maximum of the 2 March 1993 flare. The contour levels are 12.5, 25, 37.5, 50, 62.5, 75, and 87.5 percent of $I_{\rm{max}}$. BPC is a triangular cusp structure at the top of flaring loop. F1 and F2 are the flare footpoints on the solar disc.} \label{fig5}
\end{figure}

Figure~\ref{fig5} shows the 23-33 keV image recorded during the maximum of HXR emission. We see the high flaring loop above the eastern solar limb (its altitude was about 45 Mm). There was a triangular cusp structure, BPC, at the top and strong footpoint sources, F1 and F2, on the solar disc. The position of the footpoints indicates that the plane of the flaring loop was tilted from the plane of the image, {\it i.e.} that there is a significant geometrical foreshortening in the north-south dimension of the loop. We have seen in Papers I and III that the sources B and C were located at the places where the cusp structure was connected with the arcade channel.

\begin{figure}
\includegraphics[width=1\textwidth]{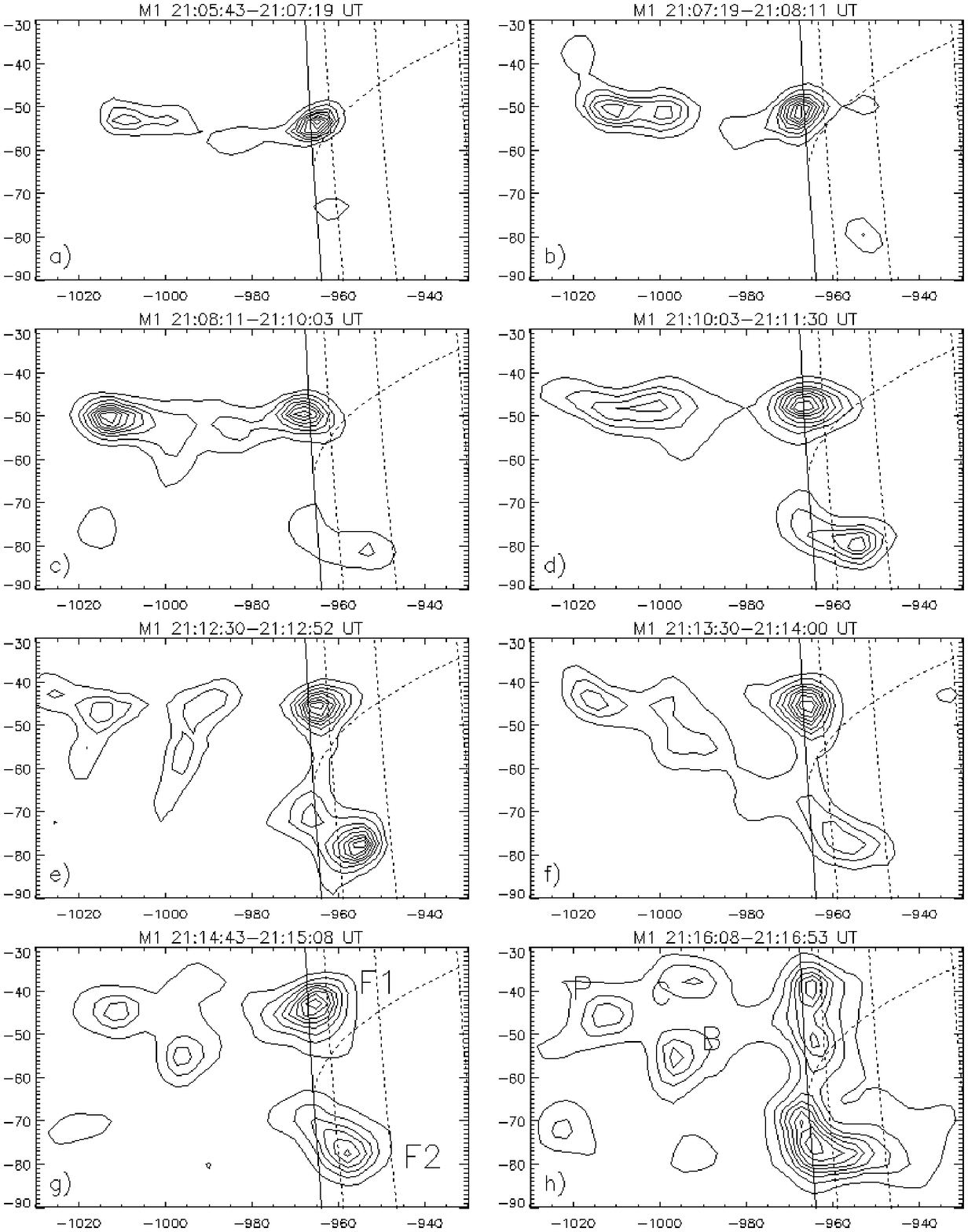}
\hfill \caption{Sequence of the 23-33 keV images of the 2 March 1993 flare. The contour levels are the same as in Figure~5.} \label{fig6}
\end{figure}

Figure~\ref{fig6} shows a sequence of 23-33 keV images. This long sequence allowed us to investigate asymmetry in electron distribution and its changes in the precipitation in some detail. Between 21:06 and 21:08 UT (Figures~\ref{fig6}a-c) there was a clear coupling between the top source P and the footpoint F1. During this time interval a gradual increase in the intensity of the P source was seen, which is certainly the result of the increase in density due to the chromospheric evaporation from the footpoint F1. About 21:10 UT (Figure~\ref{fig6}d) had begun strong precipitation toward the footpoint F2. This caused chromospheric evaporation, inflow of dense plasma into the acceleration volume, increase in the number of accelerated electrons and generation of intense HXR pulses. This moment of time ($\approx$21:10 UT) is analogous to 16:31 UT in the flare of 2 November 1991 (see Paper III) when the footpoint source F had appeared (see Figure~6 in that paper).

\begin{figure}
\includegraphics[width=1\textwidth]{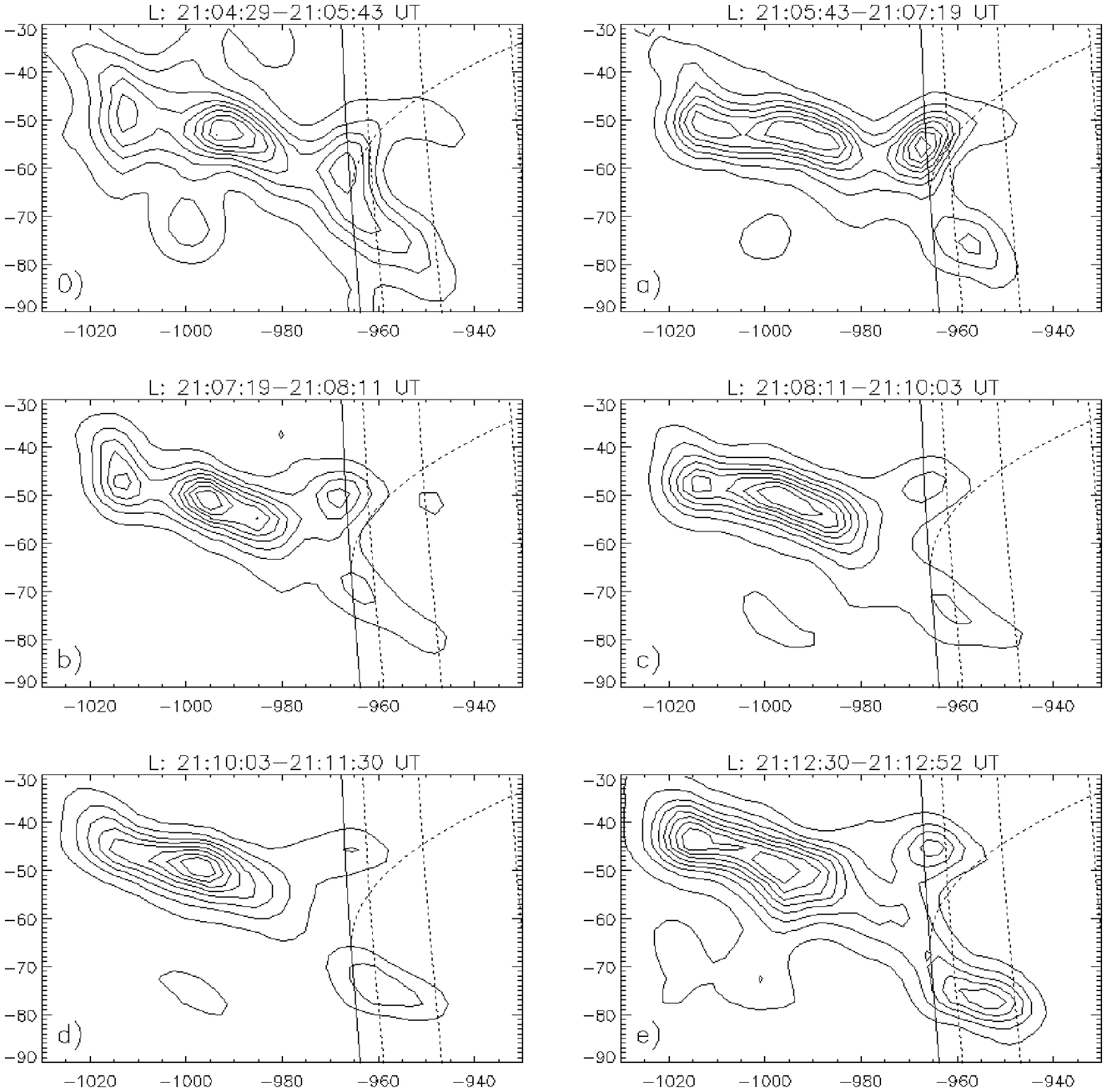}
\hfill \caption{Sequence of the 14-23 keV images of the 2 March 1993 flare.} \label{fig7}
\end{figure}

Figure~\ref{fig7} shows a sequence of the {\sl Yohkoh} HXT 14-23 keV images. In the images the loop-top sources P and B dominate. In some images (Figures \ref{fig7}d, \ref{fig7}f, and \ref{fig7}g) these sources are connected (not resolved). The footpoint sources are conspicuous only in the first three images (Figures \ref{fig7}o, \ref{fig7}a, and \ref{fig7}b) and during the HXR maximum (Figure~\ref{fig7}e). After 21:17 UT the loop-top source P dominated (Figures \ref{fig7}i, \ref{fig7}j, and \ref{fig7}k).

\begin{figure}
\includegraphics[width=1\textwidth]{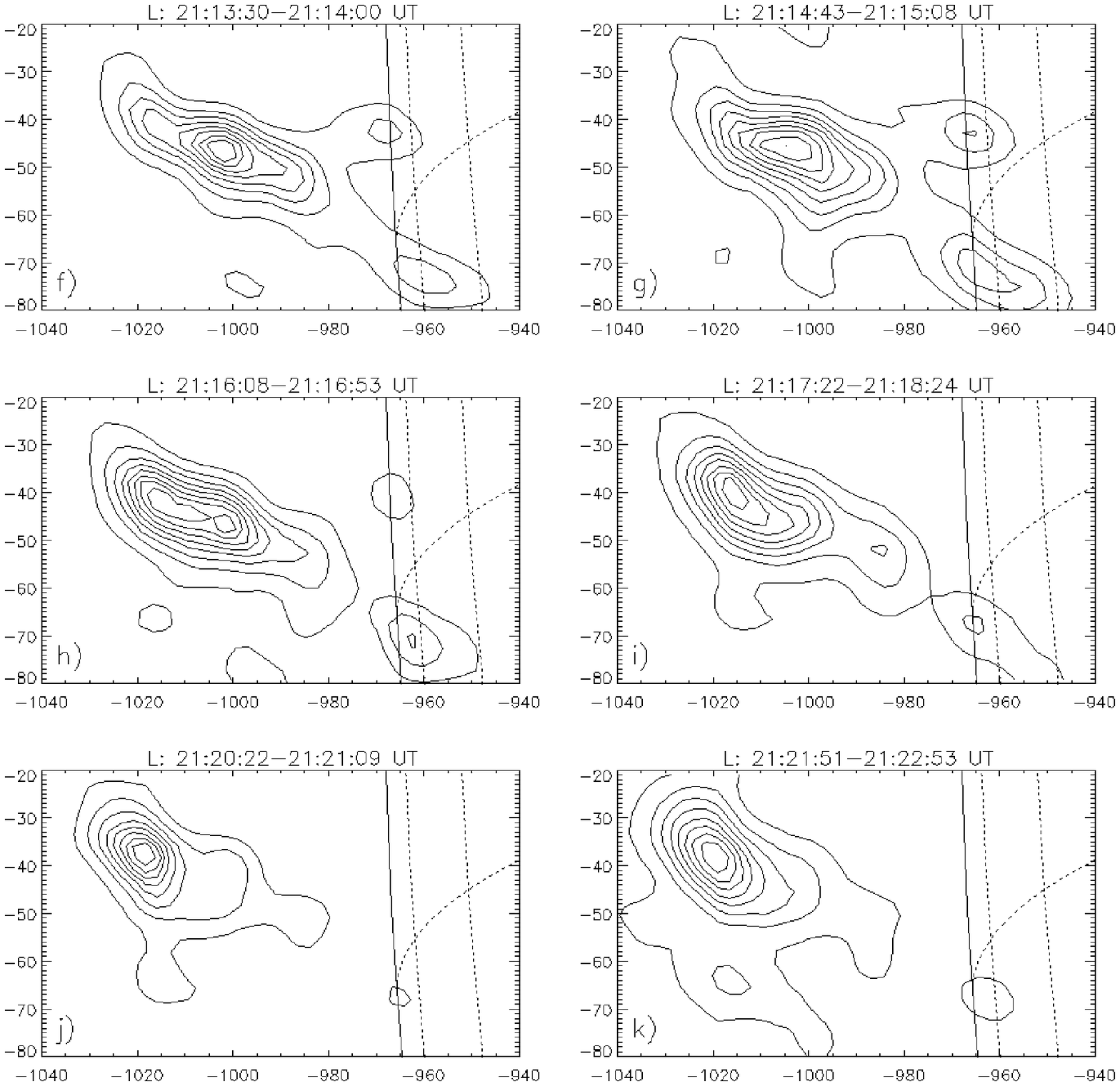}
\begin{center}
Figure~\ref{fig7} (continued)
\end{center}
\end{figure}

Figures \ref{fig6} and \ref{fig7} show comparison of significant differences in the precipitation of $\approx$15 keV and $\approx$25 keV electrons. According to our model (see Papers I-III) the cusp structure was filled with many oscillating magnetic traps. During compression of a trap its trap ratio, $\chi = B_a/B_b$ ($B_a$ is the magnetic field strength at the ends of the trap, $B_b$ is the strength at the middle of the trap), decreased and reached its minimum value, $\chi_{{\rm min}}$, during the maximum of compression. Then the electrons inside the trap reached their maximum energy and the efficiency of their precipitation reached its maximum (see Papers I-III).

The maximum of compression is different in different traps, {\it i.e.} their $\chi_{{\rm min}}$ values are different. The electrons inside the traps undergoing weaker compression (higher $\chi_{{\rm min}}$ values) achieved lower energies ($\approx$15 keV) and the efficiency of their precipitation was low (see Figure~\ref{fig7}). The electrons within the traps undergoing strong compression (low $\chi_{{\rm min}}$ values) achieved higher energies ($\approx$25 keV) and they efficiently precipitated toward the footpoints (Figure~\ref{fig6}). This means that the ensemble of the traps which was responsible for generation of most of the $\approx$15 keV electrons, was different from the ensemble providing the $\approx$25 keV electrons.

Figure~\ref{fig7} also shows that most of the $\approx$15 keV electrons emitted their energy and were thermalized within the BPC cusp structure. The strong source B in these images indicates that the traps which dominated in the generation of $\approx$15 keV electrons had good connection with the arcade channel at B. Characteristic features of the HXR impulsive phase are the asymmetry in the precipitation of accelerated electrons from the cusp structure toward the footpoints and changes in the asymmetry with time and with the energy of the electrons (see Papers I-III). These features are clearly seen in Figure~\ref{fig6} of the present paper.

During the HXR pulses 21:06-21:10 UT the precipitation toward the footpoint F1 dominated (Figures~\ref{fig6}a-c). About 21:10 UT (Figure~\ref{fig6}d) we see the development of the footpoint F2 and about 21:12 UT (Figure~\ref{fig6}e) the footpoints F1 and F2 had similar intensities. During the HXR maximum (21:13 UT, Figure~\ref{fig6}f) the precipitation toward the footpoint F1 again dominated, but during the second strong HXR pulse (21:15 UT, Figure~\ref{fig6}g) the precipitation toward the two footpoints was again of similar magnitude. In later images (after 21:16 UT, Figure~\ref{fig6}h), the precipitation toward the footpoint F2 dominated.

These complicated variations in the asymmetry of precipitation of accelerated electrons can be adequately explained in terms of our model of oscillating magnetic traps.
\begin{enumerate}[i)]
\item The asymmetry in the precipitation for an HXR pulse may arise if the axis of symmetry of the magnetic traps (approximately, it is the line joining source P with the middle of the segment BC) is not perpendicular to BC. A small deviation of the axis of symmetry from being perpendicular to BC will introduce small difference in the maximum compression (${\chi}_{\rm min}$) at the opposite ends of the traps. These small differences in the ${\chi}_{\rm min}$ values lead to significant differences in the precipitation of electrons, since the efficiency of precipitation steeply depends on ${\chi}_{\rm min}$ (see Section~3 in Paper III). Hence, large asymmetry in precipitation may result from moderate deviation of the axis of symmetry of the magnetic traps from being perpendicular to the line BC. An observation that footpoint intensities F1$>$F2 means that ${(\chi}_{\rm min})_1 < {(\chi}_{\rm min})_2$ in the cusp, F1$\approx$F2 means ${(\chi}_{\rm min})_1 = {(\chi}_{\rm min})_2$, and F1$<$F2 means ${(\chi}_{\rm min})_1 > {(\chi}_{\rm min})_2$.
\item Changes in the asymmetry of precipitation (F1/F2 ratio) mean that the direction of the axis of symmetry of the traps changes in time. Again, small changes in the direction can cause large changes in the asymmetry of precipitation, since small changes in ${\chi}_{\rm min}$ lead to large changes in precipitation.
\end{enumerate}

\begin{figure}
\centerline{\includegraphics[width=0.8\textwidth]{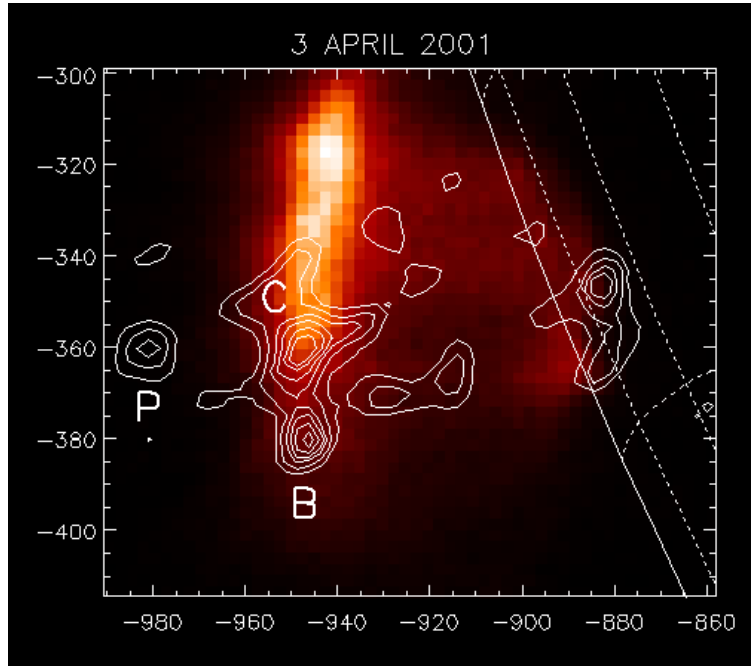}}
\hfill \caption{A {\sl Yohkoh} soft X-ray image of the 3 April 2001 flare (intensity scale in colour). The contour levels show the 23-33 keV HXR image. The solid line shows the solar limb, the dashed lines are on the solar disc.} \label{fig8}
\end{figure}

X-ray images investigated in Papers I and III have shown two important features of the large arcade flares:
\begin{enumerate}[i)]
\item The triangular (``cusp'') BPC structure was magnetically connected with the arcade channel and therefore the accelerated electrons were able to penetrate into the channel and heated it.
\item The triangular cusp covered only a part of the length of arcade channel, {\it i.e.} the channel was significantly longer than the extension of the cusp measured along the channel (see an example in Figure~\ref{fig8}). This indicates that the energy which penetrated into the arcade channel was efficiently transferred along the channel by thermal conduction. Unfortunately, the arcade channel could not be seen during the impulsive phase of the flare of 2 March 1993, since no SXR images are available for this phase.
\end{enumerate}

\subsection{Investigation of the Flare Decay Phase} \label{sec23}

\begin{figure}
\centerline{\includegraphics[width=1\textwidth]{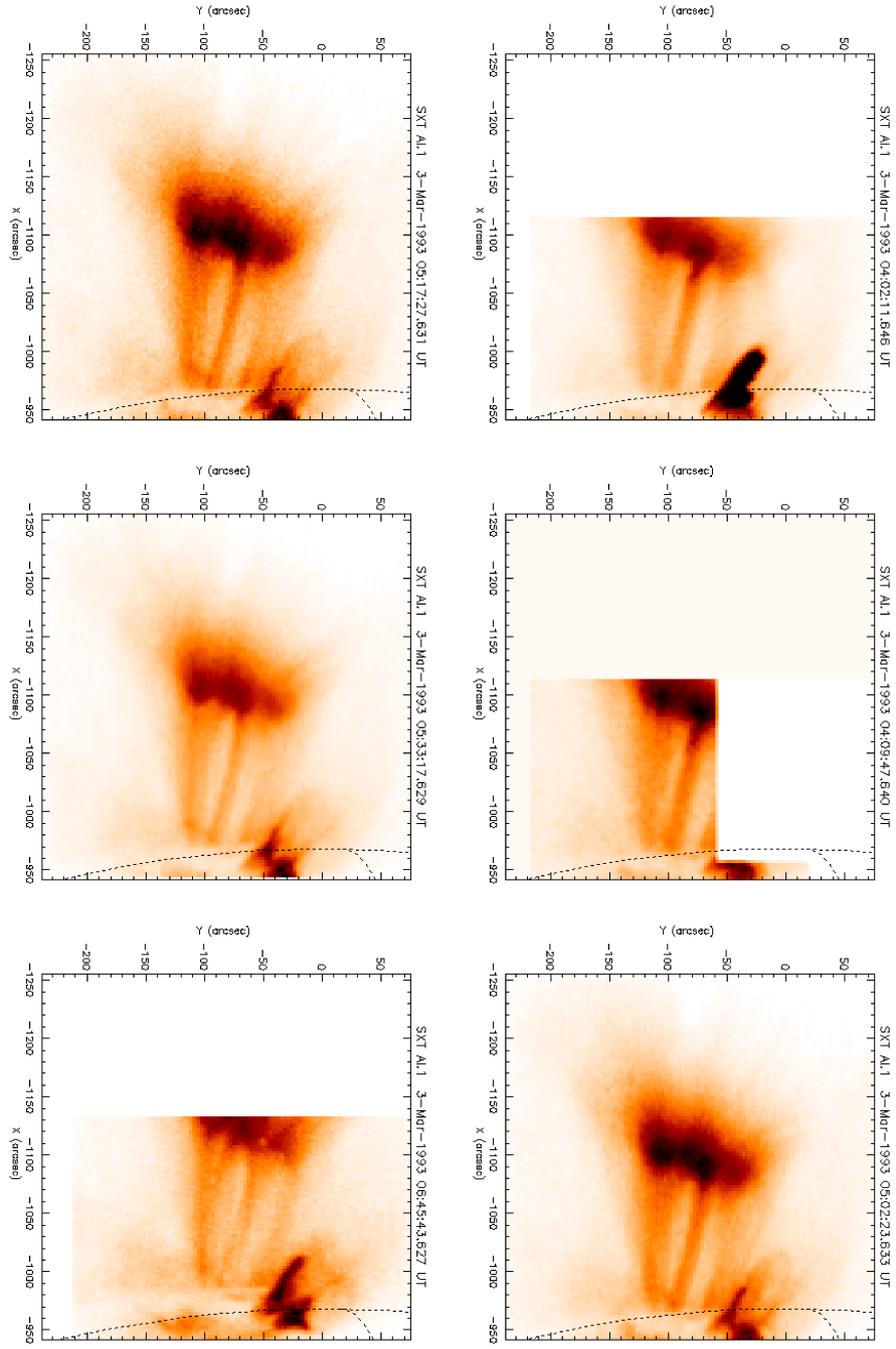}}
\hfill \caption{The {\sl Yohkoh} soft X-ray images for the late decay phase of the 2 March 1993 flare (reverse intensity scale in colour). The dashed line shows the solar limb.} \label{fig9}
\end{figure}

There were no SXR observations for the impulsive phase of the investigated flare of 2 March 1993. Some SXR images, recorded with the thin Al.1 filter, were available only for the late decay phase of the flare (Figure~\ref{fig9}). We see a long arcade of SXR loops at the altitude of about 95 Mm.

The response function of the {\sl Yohkoh} SXT observations with the Al.1 filter weakly depends on temperature in the wide range of temperatures $T \approx 2.5-20$ MK \cite{tsu91}. Therefore all plasma having $T > 2.5$ MK efficiently contributed to the recorded emission. The distribution of the intensity in the SXR images displays the distribution of the SXR emitting plasma.

\begin{figure}
\centerline{\includegraphics[width=1\textwidth]{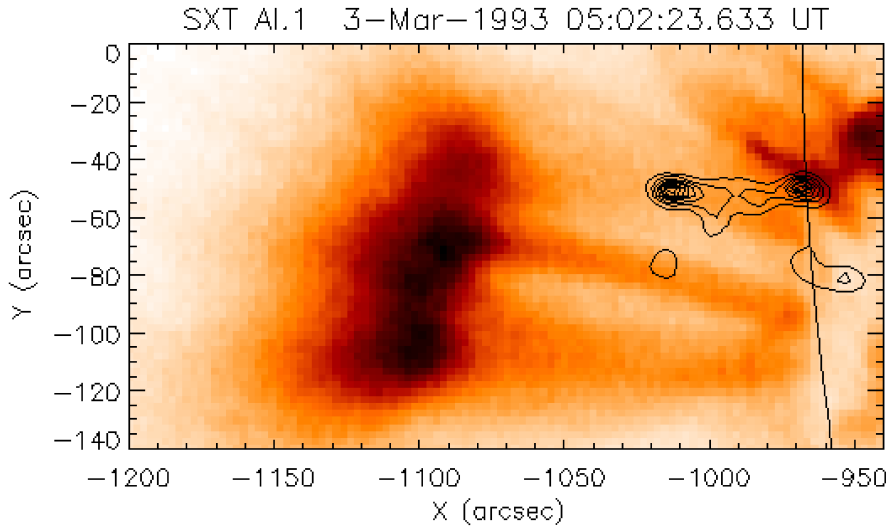}}
\hfill \caption{ The {\sl Yohkoh} soft X-ray image recorded with the Al.1 filter at 05:02:23 UT (reverse intensity scale in colour). The contour levels show the 23-33 keV HXR image recorded at 21:09 UT. See text for discussion.} \label{fig10}
\end{figure}

Figure~\ref{fig10} shows the comparison of a SXR arcade image with an impulsive-phase HXR image. The HXR image was recorded at 21:08 UT, when the arcade channel was at the level of B and C sources (see Figure~\ref{fig5}). Figure~\ref{fig10} shows that the BPC cusp structure covered only a small part of the arcade length (compare this figure with Figure~\ref{fig8}).

\begin{figure}
\centerline{\includegraphics[width=1\textwidth]{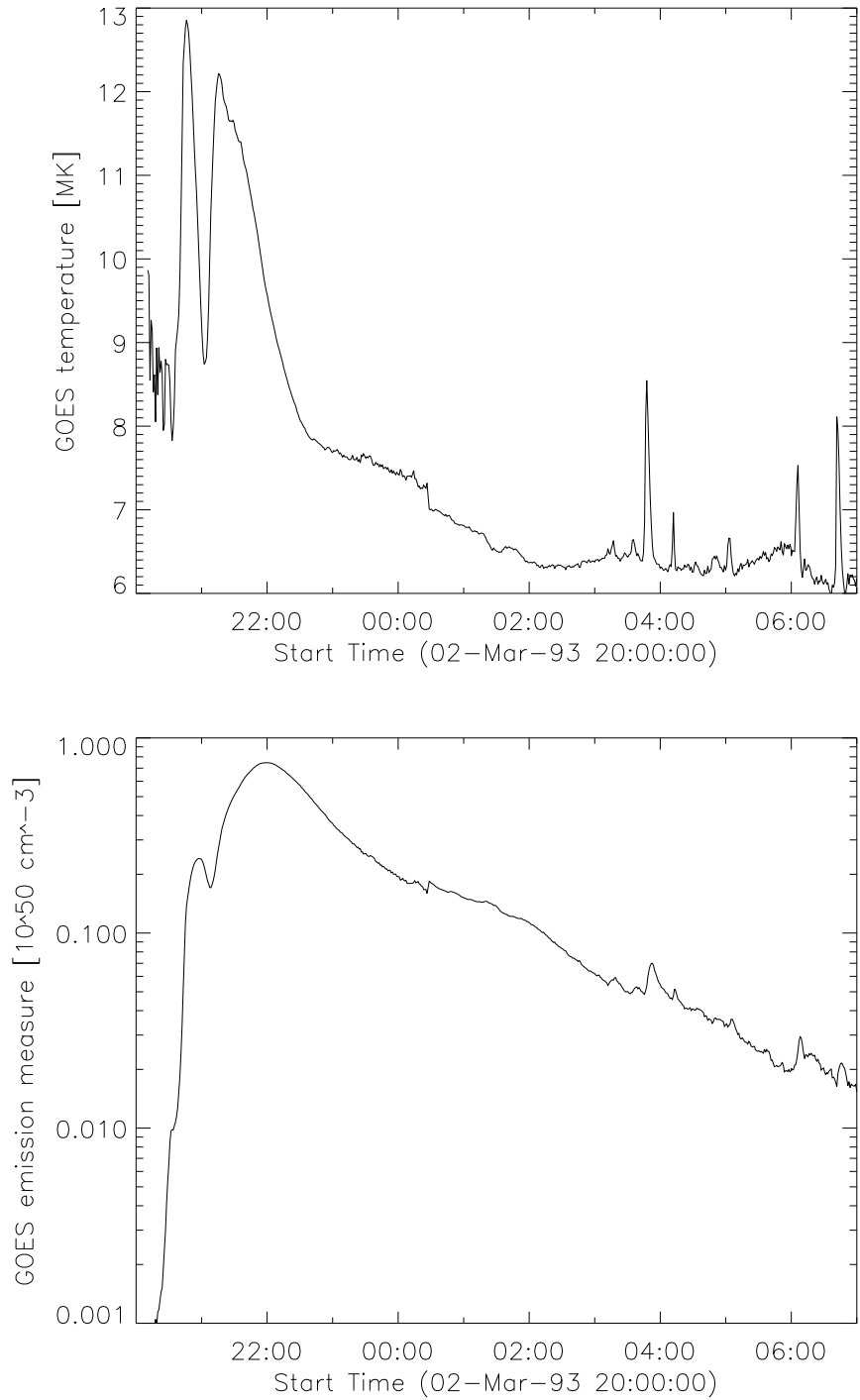}}
\hfill \caption{Time variation of the temperature (top) and the emission measure (bottom) derived from the GOES observations.} \label{fig11}
\end{figure}

In Figure~\ref{fig11} we show the time variation of the temperature, $T$, and emission measure, EM, derived from the standard GOES observations (Figure~\ref{fig2}). During the quick increase of temperature (21:00-21:20 UT; impulsive phase) the cusp structure and the arcade channel had been filled with the plasma coming from the chromospheric evaporation.

Between 21:20 and 22:00 UT the temperature decreased, but the emission measure continued to increase. This means that the energy release decreased, but it was sufficiently high to maintain the chromospheric evaporation. The decay phase of the investigated flare started at about 22:00 UT.

In steady-state loops the following condition is fulfilled:
\begin{equation}
\int E_{\rm H}\,dV = \int E_{\rm R}\,dV
\end{equation}
($E_{\rm H}$ is the heating rate per unit volume, $E_{\rm R}$ is the radiative loss per unit volume, the integrals are over the whole volume of a loop), {\it i.e.} the total heating is balanced by the total radiative loss from the loop. The ``scaling law'' of \inlinecite{ros78} is an analytical expression of the energy balance [Equation~(2)]. It can be written in the following form (see \opencite{b+j05}):
\begin{equation}
N \approx 2.1 \times 10^6\,T^2/L,
\end{equation}
where $N$ is the mean electron number density in cm$^{-3}$, $T$ is the temperature at the top of the loop in K, and $L$ is the semilength of the loop in cm.

It is very important that coronal loops have an efficient mechanism which allows them to satisfy the energy balance condition [Equation~(2)], {\it i.e.} to achieve a steady state. Here we briefly describe this mechanism. If $\int E_{\rm H}\,dV > \int E_{\rm R}\,dV$ then the flux of energy which reaches the footpoints by thermal conduction is large and it generates chromospheric evaporation. This increases density in the loop and increases $\int E_{\rm R}\,dV$ which allows the loop to achieve the energy balance [Equation~(2)]. If, on the other hand, $\int E_{\rm H}\,dV < \int E_{\rm R}\,dV$ the flux of energy which reaches the footpoints is low, since most of the energy is emitted above the footpoints. Therefore the pressure at the footpoints is too low to balance the weight of plasma contained in the loop and some amount of plasma precipitates to the chromosphere (this is seen in numerical simulations described in \inlinecite{jak92}). The density in the loop and $\int E_{\rm R}\,dV$ decrease and this allows the loop to achieve the energy balance [Equation~(2)].

This mechanism of self-regulation of coronal loops is quick in comparison with slow evolution of a flare during the slow decay phase. For example, numerical simulations have shown that a loop of semilength $L = 20$ Mm fits a change in heating rate in the time of only 5 min \cite{jak92}.

In previous papers we investigated flare evolution in ${\log}T$ vs. ${\log}N$ or ${\log}T$ vs. ${\log}\sqrt{\rm EM}$ diagnostic diagrams (see \inlinecite{j+b11} and references therein). We have found that during the slow decay phase flares evolve along the line of steady-state loops, {\it i.e.} the line described by Equation~(3) with $L = {\rm const}$. This indicates that during this evolution flares are close to the steady state [Equation~(2)] with slowly decreasing heating, {\it i.e.} decreasing values of the integrals in Equation~(2) (quasi-steady-state, or QSS, evolution of the loops). These results have been supported by numerical simulations of loops with slowly decreasing heating \cite{jak92}. These results can be summarized as follows:
\begin{enumerate}[i)]
\item During the slow decay phase (after 22:30 UT in Figure~\ref{fig11}) the loops seen in Figure~\ref{fig9} slowly evolved along a sequence of steady states (QSS evolution).
    \item The loops were continuously heated to support this QSS evolution.
\end{enumerate}

How this continuous heating of the loops seen in Figure~\ref{fig9} was performed? Our proposed explanation is the following:
\begin{enumerate}[i)]
\item Magnetic reconnection occurred at the tops of the loops and between the loops and arcade channel. This provided continuous heating of the loops.
\item New magnetic loops which were generated by the reconnection comprised only a small fraction of the volume of loops seen in Figure~\ref{fig9} and they quickly achieved the energy balance [Equation~(2)] due to the self-regulation described above. Therefore the reconnecting loops did not disturb much the slow QSS evolution of the loops. This interpretation is supported by the close correlation between the temperature and emission measure during the slow decay phase ($d\,{\log}\,T/d\,{\log}\,\sqrt{\rm EM}) \approx 0.5$ in agreement with Equation~(2) and numerical simulations described in \opencite{jak92}).
\end{enumerate}

The two main processes of thermal energy loss from a flare kernel are thermal conduction and radiative losses. The rates of these losses, calculated per unit volume, are (see \opencite{b+j05}):
\begin{equation}
E_{\rm C} \approx 3.9{\times}10^{-7}T^{3.5}/(aL) \quad\mbox{[erg\,cm$^{-3}$\,s$^{-1}$]}\quad
\end{equation}
and
\begin{equation}
E_{\rm R} = N^2{\Phi}(T) \quad\mbox{[erg\,cm$^{-3}$\,s$^{-1}$]}\quad
\end{equation}
where $a$ is the radius of the X-ray kernel, $L$ is the length of the flaring loop ``legs'' and ${\Phi}(T)$ is the radiative loss function.

When the observed temperature changes are slow, {\it i.e.} the values of $dT/dt$ are small, the loss of energy from the kernel is compensated for by the heating $E_{\rm H}$:
\begin{equation}
E_{\rm H} \approx E_{\rm C} + E_{\rm R}.
\end{equation}

We have applied Equations (3)-(6) to the beginning of the decay phase (22:30 UT in Table~\ref{tab2}) and also to the temperature maximum (21:18 UT), since the change in the heating was slow also then (small $dT/dt$ means small $dE_{\rm H}/dt$). In Figure~\ref{fig5} we have measured the radius of the cusp BPC structure, $a \approx 7.9$ Mm, and the length of the loop ``legs'', $L \approx 26$ Mm. Using Equations (3)-(6) we have calculated the values of physical parameters which are given in Table~\ref{tab2}.

\begin{table}
 \caption{Physical parameters for the time interval 21:18-22:30 UT}
 \label{tab2}
\begin{tabular}{cccccc}
\hline Time & $T$ & $N$ & $E_{\rm C}$ & $E_{\rm R}$ & $E_{\rm H}$ \\
\cline{4-6}
 [UT] & [MK] & [10$^{10}$ cm$^{-3}$] & \multicolumn{3}{c}{[erg\,cm$^{-3}$\,s$^{-1}$]} \\
 \hline
21:18 & 12.2 & 4.0$^{\ast}$ & 1.21 & 0.08 & 1.29 \\
22:30 & 8.0 & 5.2 & 0.28 & 0.14 & 0.42 \\
 \hline
\end{tabular}
\begin{list}{}{}
\item$^{\ast}$ This value has been obtained by scaling $N = 5.2 \times 10^{10}$ cm$^{-3}$ \\ (lower row) according to the change in $\sqrt{\rm EM}$, since the scaling \\ law [Equation~(3)] is not valid before the decay phase.
\end{list}
\end{table}

Table~\ref{tab2} shows that the energy release within the cusp (magnetic reconnection at the edges of the cusp, acceleration of the electrons and their thermalization) steeply decreased after the temperature maximum. Figure~\ref{fig11}a shows that after 22:00 UT the decrease in the energy release was much slower which allowed the flare to develop the long QSS decay phase.

We also see in Table~\ref{tab2} that $E_{\rm C} > E_{\rm R}$, {\it i.e.} the conductive loss of energy was larger than the radiative loss (see also \opencite{kol07}). Hence, we can use the following simple approximation in further estimates: $E_{\rm H} \approx E_{\rm C} \sim T^{3.5}$. This gives $T \sim (E_{\rm H})^{0.286}$. The small value of the power index in the last formula implies that significant changes in $E_{\rm H}$ induce only weak changes in $T$ (see Table~\ref{tab2}).

Figure~\ref{fig9} indicates that the energy release during the decay phase occurred in the arcade of SXR loops. It is most probable that the weak energy release which maintained the temperature at $T > 6$ MK was mostly due to reconnection between the arcade loops and the arcade channel. The bright arcade loops in Figure~\ref{fig9} are the places where the reconnection was most efficient.

We have applied the above Equations (3)-(6) to one of the arcade loops. We have measured the radius in the southern loop-top kernel, $a \approx 11$ Mm, and the length of the ``leg'' connecting the kernel with the loop footpoint, $L \approx 95$ Mm. We have obtained $E_{\rm H} \approx 0.023$ erg\,cm$^{-3}$\,s$^{-1}$ and $N \approx 2.0\times 10^9$ cm$^{-3}$. Hence, to maintain the long-duration decay phase, weak energy release in the arcade loops was sufficient.

Important information is contained in the time variation of the emission measure, EM$(t)$ (Figure~\ref{fig11}b). This figure shows that during the impulsive phase a large amount of hot plasma had been accumulated within the flaring system (the flaring loop and arcade channel), and during the decay phase this amount of plasma slowly and smoothly decreased (the small peak in EM$(t)$ about 00:25 UT is of instrumental origin; the peaks after 03:00 UT are due to other flares). The relationship between the physical parameters during the decay phase was the following:
\begin{equation}
E_{\rm H}(t) \hspace{0.2cm} \rightarrow \hspace{0.2cm} T(t) \hspace{0.2cm} \rightarrow \hspace{0.2cm} N(t) \hspace{0.2cm} \rightarrow \hspace{0.2cm} {\rm EM}(t)
\end{equation}
where
\begin{equation}
T \sim E_{\rm H}^{0.286}, \hspace{0.4cm} N \sim T^2, \quad\mbox{and}\quad {\rm EM} \sim N^2.
\end{equation}
(Here changes in the loop semilength, $L$, and changes in the volume, $V$, of SXR emitting plasma have been assumed to be of minor importance.)

Combination of the relationships [Equation~(8)] gives:
\begin{equation}
{\rm EM}(t) \sim [E_{\rm H}(t)]^{1.14} \quad\mbox{or}\quad E_{\rm H}(t) \sim [{\rm EM}(t)]^{0.88}.
\end{equation}
This simple relationship stresses the fact that the slow decrease in EM during the decay phase is due to slow decrease in the heating $E_{\rm H}$. In other words, the observed slow and smooth decrease of the emission measure, EM$(t)$, indicates that the thermal energy release, $E_{\rm H}(t)$, decreased slowly and smoothly during the long decay phase. It is most probable that this weak energy release was mostly due to the reconnection between the magnetic arcade loops and the arcade channel.

\section{Discussion} \label{sec3}

In Paper III we have found clear observational evidence that the strong HXR pulses at the HXR maximum (see Figure~\ref{fig1}a) were the result of inflow of dense plasma (coming from the chromospheric evaporation) into the acceleration volume inside the cusp structure. For the investigated flare of 2 March 1993 we were not able to monitor the inflow of plasma into the cusp, since we had no SXR imaging observations for the flare impulsive phase. We can, however, obtain simple estimates for the increase in density inside the cusp using only the HXR light curves.

In Table~\ref{tab1} we see that the time interval, $P_i$, between the pulses increased during the HXR maximum ($P_2/P_1 = 1.42$, where $P_1$ is the period before 21:10 UT, and $P_2$ is the period during the HXR maximum). We assume that this increase in the period is the result of increase in density and we consider two extreme cases:
\begin{enumerate}[a)]
\item We assume that $B^2/8{\pi} \gg p$ inside the traps, where $B$ is the magnetic field strength and $p$ is the pressure. Then the magnetic field does not change significantly during the increase in pressure in the traps and we have:
\begin{equation}
P_2/P_1 = (v_{\rm A})_1/(v_{\rm A})_2 = \sqrt{{\rho}_2/{\rho}_1} \approx 1.42
\end{equation}
or
\begin{equation}
{\rho}_2/{\rho}_1 = (P_2/P_1)^2 \approx 2.0,
\end{equation}
where $v_{\rm A}$ is the Alfv$\rm{\acute{e}}$n speed.
\item We assume that $B^2/8{\pi} \approx p$. Then the increase in pressure causes broadening of the traps and the magnetic field in the traps decreases. For this estimate we assumed that
\begin{equation}
B_1/B_2 = \sqrt{{\rho}_2/{\rho}_1},
\end{equation}
where $B_1$ is the magnetic field strength before the HXR maximum and $B_2$ is the field strength during the maximum. Then
\begin{equation}
(v_{\rm A})_1/(v_{\rm A})_2 = (B_1/B_2)\,\sqrt{{\rho}_2/{\rho}_1}.
\end{equation}
Combination of Equations (12) and (13) gives:
\begin{equation}
(v_{\rm A})_1/(v_{\rm A})_2 = {\rho}_2/{\rho}_1,
\end{equation}
and therefore
\begin{equation}
P_2/P_1 = (v_{\rm A})_1/(v_{\rm A})_2 = {\rho}_2/{\rho}_1 = 1.42
\end{equation}
{\it i.e.}
\begin{equation}
{\rho}_2/{\rho}_1 = 1.42.
\end{equation}
\end{enumerate}

On the other hand, we can estimate the ratio of the densities directly from the ratio, $J_2/J_1$, of the HXR fluxes ($J_1$ is the flux between 21:05-21:08 UT and $J_2$ is the flux at the HXR maximum). For the 23-33 keV flux the ratio was $J_2/J_1 \approx 2.0$ (see Figure~\ref{fig3}). During these time intervals the emission came mainly from the flare footpoints (see Figure~\ref{fig6}), hence it is proportional to the number of precipitating electrons per second which, in turn, is proportional to the electron number density, $N$, within the cusp. Hence,
\begin{equation}
N_2/N_1 \approx J_2/J_1 \approx 2.0.
\end{equation}
The last estimate is independent of the previous two. Therefore, we consider these estimates to be a confirmation that the density has quickly increased by a factor between 1.4 and 2.0 (the mean value is 1.8$\pm$0.2) and the longer period, $P_i$, at the HXR maximum is the result of this increase in density.

\section{Summary} \label{sec4}

The sequence of HXR images for the impulsive phase of 2 March 1993 flare has allowed us to investigate asymmetry in the precipitation of accelerated electrons from the acceleration volume within the cusp toward the flare footpoints. According to our model of acceleration of the electrons in oscillating magnetic traps, the precipitation was most efficient during the maximum compression of the traps (see Papers I and III). The asymmetry of precipitation shows that the maximum of compression was different at the opposite ends of the traps, {\it i.e.} the axis of symmetry of the traps was slightly inclined toward one of the ends of the traps (see Paper III). The changes in the asymmetry from one HXR pulse to another indicate that the inclination of the axis of symmetry of the traps changed.

SXR images, which were available for the late decay phase, show a long arcade of SXR loops (Figure~\ref{fig9}). In Figure~\ref{fig10} a contrast between the ``slim'' shape of the flaring loop and the long arcade is seen. This contrast is enhanced by the fact that there were no SXR images for the flare impulsive phase and therefore the arcade channel was not seen during this phase. In other flares such SXR images show a connection between the cusp and arcade channel and the heating of the channel by the cusp (see Figure~\ref{fig8}). Similar cases like in Figure~\ref{fig8} will be shown in our next paper.

Important information about the evolution of a flare during slow decay phase is contained in the time variation of the temperature, $T(t)$, and emission measure, EM$(t)$. This information is the following:
\begin{enumerate}[i)]
\item weak heating occurs during the slow decay phase and it slowly decreases;
\item the decrease in the heating determines slow and smooth decrease in EM;
\item the coupling between the heating and the amount of the hot plasma makes the flare evolve along a sequence of quasi-steady states during the slow decay phase (QSS evolution).
\end{enumerate}

\begin{acks}
This paper is dedicated to Professor Antoni Opolski on the occasion of his 100th birthday. Professor Opolski was the main person at our Astronomical Institute for many years and he efficiently stimulated the development of solar research at the Institute. We are very thankful to him for his well-wishing support. The {\sl Yohkoh} satellite is a project of the Institute of Space and Astronautical Science of Japan. The {\sl Compton Gamma Ray Observatory} is a project of NASA. The authors are very thankful to the anonymous referee for her/his important remarks which helped to improve this paper. We acknowledge financial support from the Polish National Science Centre grant 2011/03/B/ST9/00104.
\end{acks}

\end{article}

\end{document}